\documentclass[12pt,epsf]{iopart}
\usepackage{graphicx}
\usepackage{amsfonts}
\usepackage{amssymb}
\usepackage{latexsym}
\usepackage{color}
\input{colordvi.tex}
\newcommand{\CO}{{\cal O}}

\newcommand{\bear}{\begin{array}}  \newcommand{\eear}{\end{array}}
\newcommand{\bea}{\begin{eqnarray}}  \newcommand{\eea}{\end{eqnarray}}
\newcommand{\beq}{\begin{equation}}  \newcommand{\eeq}{\end{equation}}
\newcommand{\bef}{\begin{figure}}  \newcommand{\eef}{\end{figure}}
\newcommand{\bec}{\begin{center}}  \newcommand{\eec}{\end{center}}
  
\newcommand{\lmk}{\left(}  \newcommand{\rmk}{\right)}

\begin{document}

\title{Adiabatic fluctuations from cosmic strings in a contracting
universe}
\author{Robert H. Brandenberger}
%
%\affiliation
\address{Department of Physics, McGill University, 3600 University Street,
Montr\'eal QC, H3A 2T8, Canada}
\author{Tomo Takahashi}
%
%\affiliation
\address{Department of Physics, Saga University, Saga 840-8502, Japan}
\author{Masahide Yamaguchi}
%
%\affiliation
\address{Department of Physics and Mathematics, Aoyama Gakuin
University, Sagamihara 229-8558, Japan, \\ and \\
Department of Physics, Stanford University, Stanford CA 94305}
%%%

\date{\today}
%%

%\maketitle

\begin{abstract}
  We show that adiabatic, super-Hubble, and almost scale invariant
  density fluctuations are produced by cosmic strings in a contracting
  universe. An essential point is that isocurvature perturbations
  produced by topological defects such as cosmic strings on
  super-Hubble scales lead to a source term which seeds the growth
  of curvature fluctuations on these scales. Once the symmetry
  has been restored at high temperatures, the isocurvature seeds
  disappear, and the fluctuations evolve as adiabatic ones in
  the expanding phase. Thus, cosmic strings may be resurrected
  as a mechanism for generating the primordial density fluctuations 
  observed today.
\end{abstract}


\maketitle

%\newpage
%\tighten

%%%%%%%%%%%%%%%%%%%%%%%%%%%%%%%%%%%%%%%%%%%%%%%%%%%%%%%%%%%%%%%%%%%%%
\section{Introduction}
%%%%%%%%%%%%%%%%%%%%%%%%%%%%%%%%%%%%%%%%%%%%%%%%%%%%%%%%%%%%%%%%%%%%%

\label{sec:intro}

The cosmic microwave background (CMB) anisotropies observed by the
Wilkinson Microwave Anisotropy Probe (WMAP) have revealed that the
primordial density fluctuations are almost adiabatic and scale invariant
\cite{Komatsu:2008hk,Dunkley:2008ie}.  Inflationary cosmology
\cite{inflation} is currently the favored scenario to explain such
primordial fluctuations. According to it, density perturbations are
produced as quantum vacuum fluctuations on sub-Hubble scales and then
stretched to super-Hubble scales during the phase of accelerated
expansion of space. However, inflationary cosmology is not without its
problems (see e.g.  \cite{RHBrev}),\footnote{One of the problems in the
usually considered mechanism, where quantum fluctuations are the origin
of the present cosmic fluctuations, is the transition from quantum
quantities to classical observables. However, in the mechanism proposed
in this paper, such a problem does not exist because they are classical
fluctuations from the beginning.} and thus it is important to study
scenarios alternative to inflation. In the 1980s, topological defect
models such as those based on cosmic strings were investigated intensely
as a possible alternative to generate primordial density fluctuations
\cite{topologicaldefects}. However, the fluctuations induced by defects
in an expanding universe are isocurvature and, even if they might mimic
the inflationary predictions for the temperature-temperature (TT)
correlation of the CMB \cite{Turok:1996wa}, observations of the
anti-correlation of the temperature and the E-mode polarization (TE),
precisely measured by WMAP, confirmed that such fluctuations could not
be the dominant source of CMB anisotropies
\cite{Peiris:2003ff,Komatsu:2008hk}. Thus, causal scaling seed models
are ruled out as a main component of primordial density fluctuations.

On the other hand, in recent years other types of scenarios alternative
to inflation motivated by developments in string theory have been
proposed. Examples are the Pre-Big-Bang model (PBB)
\cite{Gasperini:1992em} and the Cyclic/Ekpyrotic scenario
\cite{Khoury:2001wf,Steinhardt:2001st}.\footnote{Another model is string
gas cosmology \cite{BV,NBV}, in which even the dimensionality of the
observable universe may be determined dynamically.}  The common feature
of these models is that the universe begins in a contracting phase
before emerging into the expanding phase of Standard Big Bang cosmology
after a cosmological bounce. In the contracting phase, comoving scales
exit the Hubble radius unless the contraction is too rapid. Previous
studies have considered quantum mechanical vacuum fluctuations of a
scalar matter field evaluated when the matter scales exit the Hubble
radius during the contracting phase. Of course, once they are produced
in the contracting phase, the fluctuations must be coupled to
fluctuations in the expanding phase after the bounce. The propagation of
fluctuations through the bounce phase depends on the details of the
bounce \cite{Durrer}.  There are models which yield an almost scale
invariant spectrum (see e.g. \cite{Tolley}) after the bounce.

In this paper, we suggest the possibility that primordial density
fluctuations are produced by causal seeds such as cosmic strings in the
contracting phase, and show that they could generate adiabatic, almost
scale invariant, and super-Hubble curvature fluctuations in the
expanding universe.\footnote{Another attempt to produce adiabatic
fluctuations from cosmic strings is discussed in the context of
two-metric theories of gravity \cite{Avelino:2000iy}.} One simple
possibility to realize cosmic strings in the contracting universe is to
embed our model into a so-called cyclic universe, in which cosmic
strings are formed in the usual way during a phase transition in the
expanding phase, if the matter Lagrangian admits cosmic strings. In our
scenario, different from the cyclic scenario of Steinhardt and Turok
\cite{Steinhardt:2001st} where quantum fluctuations source the
primordial density fluctuations, here cosmic strings seed the
perturbations. Of course, in the cyclic scenario, topological defects
may be dangerous because they may dominate the energy density of the
universe, as pointed out by Avelino et al. in
Refs.~\cite{Avelino:2002hx,Avelino:2002xy}. However, Avelino et al. also
give a solution to that problem: they point out that a relatively long
period of cosmic acceleration at low energies (late period of one cycle)
can dilute topological defects in order that they do not overdominate
the universe. A second possibility is to consider the birth of the
universe in the contracting universe (not cyclic). If the universe has
finite birth time, the correlated region at the birth of the universe
does not necessarily cover the whole volume of the universe. Then, the
randomness of the values of the underling field beyond the correlation
length leads to the formation of topological defects.

Density fluctuations produced by causal seed models naturally become
super-Hubble in the contracting phase. More specifically, the key point
is that the defect-seeded perturbations which are initially isocurvature
in nature seed a growing adiabatic mode. At the time when the symmetry
(whose breaking yields the topological defects) is restored, the seed
term disappears and the fluctuations become frozen-in super-Hubble scale
adiabatic perturbations. As long as no dominant isocurvature
fluctuations are produced in the expanding phase, the fluctuations in
the expanding phase will be adiabatic and thus able to explain the TE
anti-correlation observed in the CMB. In the following we consider
cosmic strings as a concrete example and investigate the nature of the
density fluctuations in detail.

%%%%%%%%%%%%%%%%%%%%%%%%%%%%%%%%%%%%%%%%%%%%%%%%%%%%%%%%%%%%%%%%%%%%%
\section{Adiabatic fluctuations from cosmic strings in a contracting
universe}
%%%%%%%%%%%%%%%%%%%%%%%%%%%%%%%%%%%%%%%%%%%%%%%%%%%%%%%%%%%%%%%%%%%%%

The evolution of cosmic strings in a contracting universe was
investigated in Refs. \cite{Avelino:2002hx,Avelino:2002xy}. As in these
references, we will assume that the distribution of strings on
super-Hubble scale is like a random walk. We make the simplest
assumption that the universe is initially matter and then radiation
dominated in the contracting phase. In this context, it was shown that
cosmic strings obey the scaling solution asymptotically both in the
radiation and matter dominated epochs.  Specifically, the correlation
length $L$ is proportional to $a^{2} \ln a \propto (-t) \ln (-t)$ in the
radiation era ($a$ is the scale factor and $t(<0)$ is cosmic time, with
the bounce time taken as $t=0$). If we take the string loop chopping
efficiency $\tilde{c}$ to be a non-zero constant, then the ratio of the
energy density in cosmic strings to the total one stays almost
constant.\footnote{There remains the logarithmic dependence in the
radiation dominated era, which can lead to a small deviation from scale
invariance of the final curvature perturbations.}  Therefore, the
density fluctuations produced by cosmic strings are almost scale
invariant at least initially at the Hubble radius exit.

On super-Hubble scales, the dynamics of the defect network
sets up isocurvature fluctuations which in turn act as a
seed for growing curvature perturbations.
As the universe contracts, the temperature of radiation
increases, and eventually leads to symmetry 
restoration and the disappearance of the
topological defects. Thus, there will no longer
be any isocurvature fluctuations, and the source on
super-Hubble scales in the differential equation for the
adiabatic mode vanishes. Hence, the fluctuations become
frozen in as adiabatic ones.

On super-Hubble scales, the equation for the evolution of the curvature
perturbation on uniform total density hypersurfaces, $\zeta$, is given
by\footnote{
  Although the density fluctuations of cosmic strings can be large and
  of the order of unity, the linear perturbation theory applies
  because the metric perturbations remain small as long as the energy
  density of the cosmic string is subdominant.
}  \cite{Wands:2000dp}
\beq
 \dot{\zeta} = - \frac{H}{\rho+P} \delta P_{\rm nad},
 \label{eq:zetaevo}
\eeq
where $H \equiv \dot{a}/a$ is the Hubble parameter, $\rho$ and $P$ are
the total energy density and pressure, respectively. A dot represents
a derivative with respect to the cosmic time.  The non-adiabatic pressure
perturbation $\delta P_{\rm nad}$ is defined as $\delta P_{\rm nad}
\equiv \delta P - c_s^2 \delta \rho$ with $\delta \rho$ being the
total density fluctuation and $c_s^2 = \dot{P}/\dot{\rho}$ being the
adiabatic sound speed. In the multi-fluid case, the total
non-adiabatic pressure perturbation consists of two parts. The first
part comes from intrinsic entropy perturbations which vanish for a
barotropic fluid.  Therefore, the intrinsic entropy perturbations of
matter and radiation vanish. On the other hand, it is non-trivial that
the intrinsic isocurvature perturbation for cosmic strings also
vanishes as strings can be modeled as a simple fluid with equation of state 
$w_{st} \equiv P_{st} / \rho_{st} = - 1/3$. However, we expect that 
the intrinsic entropy perturbation of cosmic strings is negligible and
will assume this in the following.  The second part $\delta P_{\rm rel}$ 
comes from the relative entropy perturbation between different fluids
$S_{\alpha\beta}$ \cite{Malik:2002jb},
\beq
  \delta P_{\rm rel} \equiv \frac{1}{6H\dot{\rho}} \sum_{\alpha,\beta}
     \dot{\rho}_{\alpha} \dot{\rho}_{\beta} (c_{\alpha}^2-c_{\beta}^2)  
     S_{\alpha\beta},
  \label{eq:relnonad}
\eeq
where the relative entropy perturbation between different fluids
$S_{\alpha\beta}$ is given by
\beq
  S_{\alpha\beta} = - 3H 
           \lmk  \frac{\delta\rho_{\alpha}}{\dot{\rho_{\alpha}}}
               - \frac{\delta\rho_{\beta}}{\dot{\rho_{\beta}}} \rmk,
  \label{eq:relentropy}
\eeq  
and the adiabatic sound speed for each component, $c_\alpha^2$, is
given by $\dot{P_\alpha}/\dot{\rho_\alpha}$ with $\rho_\alpha$ and
$P_\alpha$ being the energy density and pressure of the component.

Now, let us estimate the amplitude of the curvature perturbation $\zeta$
for a comoving scale $k$. First, we consider a scale $k$ which exits the
Hubble radius during a radiation dominated era. We neglect the curvature
perturbations which are generated when the corresponding scale is
sub-Hubble.\footnote{This can be justified by assuming that string loop
distribution is subdominant and the initial value of the curvature
perturbation at $t \rightarrow - \infty$ vanishes. In fact, we expect
that the string loop distribution will be subdominant in a contracting
universe compared to an expanding universe since comoving scales are
exiting rather than entering the Hubble radius. Loops exit the Hubble
radius before they collapse through emission of gravitational
waves. This effect reduces the loop chopping efficiency $\tilde{c}$, and
hence the number of produced loops is smaller in a contracting phase.}
Thus, we just follow the evolution of the curvature perturbation from
the epoch $t_H(k)$ when a comoving scale $k$ exits the Hubble radius
until the time $t_{cs}$ when the symmetry is restored and the strings
disappear.
%Since today's adiabatic fluctuations
%originate to isocurvature ones during a contracting phase, first we
%evaluate the amplitude of the isocurvature fluctuations.  

The relative entropy perturbation between
radiation and cosmic strings is 
\beq
  S_{rs} \simeq - 3H \frac{\delta \rho_{\rm st}}{\dot{\rho}_{\rm st}},
  \label{eq:entropyrs}
\eeq
where we have used the fact that 
$\dot{\rho}_{\rm st}/\dot{\rho}_{\rm rad}$ 
is almost constant and much smaller than unity, and 
$|\delta \rho_{\rm rad}|$ is at most comparable to 
$|\delta \rho_{\rm st}|$. From the scaling solution, it follows that 
cosmic strings can be modeled as a random walk on scales
larger than the Hubble radius with step length comparable to
the Hubble radius. Then, the density fluctuations of cosmic strings
for a super-Hubble scale can be easily estimated as
\beq
  \delta \rho_{st}(k) \simeq N |H| \mu k_{\rm phys},
  \label{eq:deltarhosr}
\eeq
where $\mu$ is the mass per unit length of a string, $N=\CO(1)$ is
%an almost constant number given by $\sqrt{N} = L |H|$.  
the number of long strings crossing any given Hubble volume, and $k_{\rm
phys} \equiv k/a$. Notice that $N$ can be different in radiation and
matter dominated epochs although the difference is expected to be at
most $\CO(1)$.  Inserting these equations to Eq. (\ref{eq:zetaevo})
yields
\beq
 \dot{\zeta} \simeq N G \mu k_{\rm phys}.
\eeq

As stated above, cosmic strings disappear at the time
$t_{\rm cs}$ due to symmetry restoration. Once
cosmic strings disappear, the curvature perturbation is conserved at
least until the bounce. Thus, the final curvature perturbation before
the bounce is estimated as
\beq
  \zeta = \int_{t_H(k)}^{t_{\rm cs}} dt \dot{\zeta}
        \sim  N_{\rm r} G \mu k (-t_H(k))^{\frac 12} 
        \sim N_{\rm r} G \mu.
\eeq
Here we have made use of $k (-t_H(k))^{\frac 12} = 1/2$ (in the
radiation era), and $N_{\rm r}$ is the value of $N$ (in the
radiation epoch) which is $\CO(1)$. Thus,
the curvature perturbations are independent of the comoving scale $k$ and
hence scale invariant at least before the bounce. In the same way, the
curvature perturbation for comoving scales whose physical scales exit
the Hubble radius during the matter era is estimated as
\beq
  \zeta = \int_{t_H(k)}^{t_{\rm eq}} dt \dot{\zeta} +
          \int_{t_{\rm eq}}^{t_{\rm cs}} dt \dot{\zeta}
        \sim  N_{\rm m} G \mu k (-t_H(k))^{\frac 13}   
        \sim  N_{\rm m} G \mu,
\eeq
where $t_{\rm eq}$ is the matter-radiation equality time in the
contracting phase, we have made use of $k (-t_H(k))^{\frac 13}
= 2/3$ (in the matter era), and $N_{\rm m}$ is the value of $N$ in
the matter epoch.  Therefore, the curvature perturbation is
scale invariant at least before the bounce also on these scales.  

According to the often-used Hwang-Vishniac \cite{HV} (Deruelle-Mukhanov
\cite{DM}) matching conditions for fluctuations across a space-like
hypersurface, the curvature perturbation is conserved across the
bounce. If we apply these matching condition, we conclude that the final
curvature perturbations in the expanding phase are almost scale
invariant and hence could be responsible for the present density
fluctuations if $G \mu \sim 10^{-5}$. As emphasized in \cite{Durrer},
there are problems with blindly applying these matching
conditions. Subsequent studies have shown that the actual transfer of
the fluctuations depends quite sensitively on the details of the
bounce. There are cases where the curvature perturbation is conserved
(see e.g. \cite{Tsujikawa:2002qc,Copeland:2006tn,Bozza}), but there are
other examples where this does not hold
\cite{Hassan,Tirtho,Cai}. However, if the bounce time is short compared
to the time scale of the fluctuations of interest, it can be rather
rigorously shown that the spectrum of $\zeta$ is maintained through the
bounce. This can be shown \cite{Cai,Cai2} by modeling the background
cosmology with three phases: the initial contracting radiation phase,
the ``bounce phase'' during which $H = \alpha t$, where $\alpha$ is some
constant, and the expanding radiation phase. The matching conditions at
the two hypersurfaces between these phases can be consistently applied
(since the background also satisfies the matching conditions, unlike
what happens in the single matching between contracting and expanding
phase which has been applied in the case of the singular Ekpyrotic
bounce).
%Clearly, this topic needs to be examined in more detail. 
However, we would like to point out that, even if the
curvature perturbation is {\it not} conserved through the bounce, the
scale invariance of the final curvature perturbation still holds true as
long as its change in the amplitude of the fluctuations across the
bounce is independent of the comoving scale. This can be reasonably
expected for modes we are interested in because their momenta are much
smaller than the maximal value of $|H|$ around the bounce point
(assuming that the bounce is smooth), the only energy scale which can be
set by the bounce.\footnote{
  Even in the case when the change through the bounce depends on the
  comoving scale, a scale invariant spectrum may be realized by
  considering the time varying tension of cosmic strings discussed in
  Refs. \cite{Yamaguchi:2005gp,Ichikawa:2006rw,Takahashi:2006yc}, which
  compensates for the variation of the curvature perturbation.
}

Finally, we comment on some subtleties. First of all, cosmic strings may
be formed again in the expanding phase. In this case, cosmic strings
again produce isocurvature fluctuations in the expanding phase, which
should be suppressed to less than $10\%$ of the total curvature
perturbations \cite{Pogosian:2003mz,Endo:2003fr}. Such a suppression may
be realized in the case when the constant $N$ for cosmic strings in the
contracting phase is much larger than that in the expanding phase. The
fact that the loop chopping efficiency $\tilde{c}$ is smaller implies
that the constant $N$ might be larger. Therefore, it is plausible that
the constant $N$ for cosmic strings in the contracting phase is larger
than that in the expanding phase. Another possibility is that the
curvature perturbation sourced by fluctuations of cosmic strings in a
contracting phase is amplified at the bounce. Another solution is to
simply assume that cosmic string are not produced in the expanding phase
because the symmetry breaking patterns of scalar fields are not
necessarily the same in the contracting and the expanding phases.

Another issue is that cosmic strings generate not only density
perturbations but also vector and tensor perturbations (gravitational
waves). Gravitational waves are produced by oscillations of loops as
well as long strings. As explained before, the radius of a loop could
become larger than the Hubble before there has been a significant amount
of gravitational radiation. Therefore, we expect that the relative
amplitude of gravitational waves to scalar metric fluctuations will be
smaller in the contracting phase than the standard scenario of cosmic
strings in an expanding universe.  Regarding the vector mode, it has
been shown that vector perturbations exhibit growing mode solutions in
the contracting phase \cite{Battefeld:2004cd}. In particular, the metric
perturbations always grow while the matter perturbations stay constant
in the radiation dominated era. However, the vector perturbations will
decrease in the expanding phase. If vector fluctuations are suppressed
(even slightly) across the bounce (or the scalar fluctuations enhanced),
then the vector modes will be sufficiently small today not to destroy
the successful agreement with the CMB angular power spectrum. Small but
not negligible vector fields could, in fact, be useful for generating
the observed large scale magnetic fields, as pointed out in
Ref. \cite{Battefeld:2004cd}. As a final remark, we mention possible
non-Gaussianities in the scenario. Since the distribution of cosmic
string is highly non-Gaussian, the produced density fluctuations may
give large non-Gaussianity, though the bispectrum for a simulated string
model in the expanding universe is shown not to be so large for the
diagonal contribution \cite{Gangui:2001fr}. All these topics are worth
investigating.

%%%%%%%%%%%%%%%%%%%%%%%%%%%%%%%%%%%%%%%%%%%%%%%%%%%%%%%%%%%%%%%%%%%%%
\section{Summary}
%%%%%%%%%%%%%%%%%%%%%%%%%%%%%%%%%%%%%%%%%%%%%%%%%%%%%%%%%%%%%%%%%%%%%

We have shown that adiabatic, super-Hubble, and almost scale invariant
density fluctuations can be produced by cosmic strings in a contracting
universe. Although cosmic strings can only generate isocurvature
fluctuations in an expanding universe, they can produce adiabatic
fluctuations by considering them in a contracting universe because
cosmic strings disappear due to symmetry restoration. Our findings open
the possibility that topological defects could be resurrected as the
main source of the current cosmic density fluctuations.

%\bigskip

\ack
%\acknowledgments{
M.Y is grateful to M. Kawasaki for useful
  discussions. T.T. and M.Y would like to thank R. H. Brandenberger for
  kind hospitality at McGill University where this work was finished.
  This work is supported in part by a Canadian NSERC Discovery Grant and
  by the Canada Research Chair program (R.B.), and by the Sumitomo
  Foundation (T.T.), and by Grant-in-Aid for Scientific Research from
  the Ministry of Education, Science, Sports, and Culture of Japan
  No.\,19740145 (T.T.), No.\,18740157, and No.\,19340054 (M.Y.).
%}

\end{document}